Title: **Laser induced ultrafast 3d and 4f spin dynamics in CoDy ferrimagnetic alloys as a function of temperature.**

T. Ferté[1], G. Malinowski[2], E. Terrier[1], V. Halté[1], L. Le Guyader[3], K. Holldack[3], M. Hehn[2], C. Boeglin[1] and N. Bergeard[1*]

[1] *Université de Strasbourg, CNRS, Institut de Physique et Chimie des Matériaux de Strasbourg, UMR 7504, F-67000 Strasbourg, France.*

[2] *Institut Jean Lamour, Université de Lorraine, BP 50840, 54011 Nancy, France*

[3] *Institut für Methoden und Instrumentierung der Forschung mit Synchrotronstrahlung Helmholtz-Zentrum Berlin für Materialien und Energie GmbH, Albert-Einstein-Str. 15, 12489 Berlin, Germany*

*<u>Corresponding author:</u>*
*Mel: nicolas.bergeard@ipcms.unistra.fr*
*Address:* Institut de Physique et de Chimie des Matériaux de Strasbourg (IPCMS)
Campus Cronenbourg 23 rue du Loess BP43 67034 Strasbourg

0. Abstract

We report an element- and time-resolved investigation of femtosecond laser induced ultrafast dynamics of the Co 3d and Dy 4f spins in a ferrimagnetic $Co_{80}Dy_{20}$ alloy as a function of the temperature. We observe that the Co characteristic demagnetization time ($\tau_{Co}$) remains nearly constant (~0.2 ps) on increasing the temperature. Conversely, the Dy characteristic demagnetization time ($\tau_{Dy}$) decreases from ~1 ps to ~0.4 ps with the rise of temperature. Comparing our experimental data with literature shows that $\tau_{Co}$ and $\tau_{Dy}$ are independent of the alloy composition or the demagnetization amplitude and that $\tau_{Dy}$ scales with the relative temperature $T^* = T_{Curie} - T$.

*<u>Key words</u>*:

**Ferrimagnetic alloys, ultrafast laser induced demagnetization, femtosecond laser, time-resolved X-ray Magnetic Circular Dichroism**.

*1. Introduction*

Excitation of ferromagnetic layers with infra-red (IR) femtosecond (fs) laser pulses leads to a quenching of the magnetic order on a sub-picosecond time scale [1]. The microscopic mechanisms governing this ultrafast demagnetization are still subject to controversy in spite of intensive experimental and theoretical studies [2 - 8]. Although these investigations have revealed fundamental dissimilarities in the laser induced ultrafast magnetization dynamics in transition metals (TM) [1, 9 - 11] and in rare-earth (RE) films [12 - 16], a few general features were established. For instance, a rise in temperature results in an increase of the characteristic quenching times ($\tau$) of the 3d magnetic order in pure transition metals (TM) and 5d magnetic order in pure rare-earth (RE) films [11, 13, 17]. In the specific case of pure RE layers, such behaviour is also expected for the localized 4f spins as theoretically predicted [13] and sustained by experiments that have evidenced concomitant laser induced dynamics of the 5d and 4f spins [12 - 14, 16, 18]. In these experiments, the IR fs laser pulses excite the RE 5d spins while the RE 4f spin dynamics is indirectly triggered via the 5d-4f intra-atomic exchange coupling [12, 14].

In RE-TM ferrimagnetic alloys, the RE 4f spin order originates from the Ruderman-Kittel-Kasuya-Yosida (RKKY) exchange coupling [19]. The RE 4f – RE 4f indirect exchange coupling is mediated by the RE 5d as well as the TM 3d electrons in the conduction band [20]. Interestingly, the published element- and time-resolved experiments have reported distinct laser induced ultrafast dynamics for the TM 3d and RE 4f spins in spite of this RKKY exchange coupling [21 – 30]. Recently, the Landau-Lifshitz-Bloch (LLB) model [13, 17] was extended to treat the laser induced ultrafast dynamics in multi-sublattices ferrimagnetic alloys [31] such as FeCoGd [32], CoTb [33] and FeTb [34] alloys. The calculations based on this modified LLB model have shown that the characteristic demagnetization times of both the FeCo ($\tau_{FeCo}$) and the Gd ($\tau_{Gd}$) sublattices in FeCoGd alloys strongly depend on the difference between the initial temperature (T) and the Curie temperature ($T_{Curie}$) of the alloy [32]. This theoretical work highlights explicitly the differences between the ultrafast laser induced TM 3d and the RE 4f spin dynamics in these alloys with the variation of the temperature [32]. Hennecke et al. have invoked the effect of temperature to explain the short Gd demagnetization time they have evidenced in a FeCoGd alloy [30]. Ferté et al. have earlier investigated the laser induced demagnetization in $Co_{80}Dy_{20}$ and $Co_{78}Dy_{22}$ alloys at different initial temperatures but ensuring a constant relative temperature $T_{Curie} - T$ [27]. They have shown that the response of the Dy (resp. Co) magnetization to laser excitation was the same for both alloys in line with the LLB calculations. The distinct dynamics of 3d and 4f spins is believed to be the key ingredient for ultrafast all optical switching [21, 35] as well as ultrafast spin-transfer torque assisted switching [36] in RE-TM alloys. Thus, it is of paramount importance to determine the correlation between the characteristic demagnetization times and physical parameters such as the temperature. However, systematic element- and time-resolved investigations of laser induced spin dynamics in a single RE-TM alloy as a function of temperature are still lacking.

In this work, we have studied the laser induced ultrafast dynamics of Co3d and Dy4f spins in a $Co_{80}Dy_{20}$ ferrimagnetic alloy as a function of temperature by mean of time-resolved X-Ray Magnetic Circular Dichroism (tr-XMCD) [37]. Interestingly, we report experimental evidences that the dependence of the laser induced dynamics of Co 3d and Dy 4f spins on temperature are clearly different. We observe that $\tau_{Co}$ remains nearly constant (~0.2 ps) while $\tau_{Dy}$ decreases from ~1 ps to ~0.4 ps when the temperature rises from 160K to 350K. Furthermore, a comparison of our experimental data with existing data from literature on laser induced ultrafast dynamics of Dy 4f spins in CoDy alloys [25, 27] suggests that $\tau_{Dy}$ is determined by $T^* = T_{Curie} - T$.

*2. Material and methods*

The 18 nm thick $Co_{80}Dy_{20}$ alloy layer was deposited by DC magnetron sputtering on a "heat sink" Ta(3)/Cu(20)/Ta(3) multilayer sustained by a $Si_3N_4$ membrane. The alloy was capped with a Al(3)/Ta(3) bi-layer to prevent oxidation. The $Co_{80}Dy_{20}$ alloy displays an out-of-plane magnetic anisotropy. We have recorded hysteresis loops at various temperatures using SQUID magnetometry to extract the dependence of the coercive field ($H_C$) on temperature. For these measurements, we have used a test $Co_{80}Dy_{20}$ alloy layer deposited simultaneously with the one used for the time-resolved experiments but on a Si substrate (figure 1). This figure shows a divergence of $H_C$ in the vicinity of T ~ 250K indicating the temperature of magnetic compensation [38, 39].

The tr-XMCD experiments were carried out at the femtoslicing beam line of the BESSY II synchrotron radiation source at the Helmholtz-Zentrum Berlin [37]. The magnetization dynamics have been measured by monitoring the transmission of circularly polarized X-ray pulses tuned to specific core level absorption edges as a function of a pump-probe delay for two opposite directions of the magnetic field. The photon energy was set to the $CoL_3$ and the $DyM_5$ edges using a reflection zone plate monochromator on UE56/1-ZPM. The full width at half maximum (FWHM) of the 800nm pump laser was set to 500 µm to ensure homogeneous pumping over the probed area of the sample (FWHM ~ 200 µm). A magnetic field of ±0.55 T was applied along the propagation axis of both the IR laser and the X-ray beam during the experiment. The measurements were carried out at **T\*** = 350 K (configuration 1), 400 K (configuration 2) and 540 K (configuration 3) with **T\*** = $T_{Curie} - (T_{cryo} + \Delta T)$ as illustrated in figure 2. Here, $T_{cryo}$ is the temperature of the cryostat and $\Delta T$ is the temperature elevation due to the continuous laser heating (table 1). The Curie temperature of the $Co_{80}Dy_{20}$ alloy ($T_{Curie}$ = 700 K) is extrapolated from literature [40] and from mean field calculations [41, 42]. As only one single sample was used in this experiment, any small error on the estimated value of $T_{Curie}$ would shift T\* without affecting our conclusions. We have determined that the coercive field of the $Co_{80}Dy_{20}$ alloy was below 0.55 T either below 160 K or above 300 K by monitoring the XMCD amplitude as a function of temperature. The divergence of $H_C$ between 160 K and 300 K is related to $T_{comp}$ ~250 K as illustrated in figure 1. As a consequence, the experimental parameters, such as the pump laser powers (P) and $T_{cryo}$, have been chosen so that $T_{cryo} + \Delta T$ stay in the temperature ranges that allow for magnetic saturation of our alloy. We have relied on the thermal variation of the coercive field to determine $\Delta T$ for the two different pump laser powers (P = 17 and 50 mW) that we have used during the experiment. To do so, we initially set $T_{cryo}$ = 80 K and turned on the laser. P = 17 mW was the largest laser power for which $T_{cryo} + \Delta T$ stays below T = 160K. Above this temperature, the CoDy alloy could not be saturated. As a consequence, we estimated that $\Delta T$ ~ 80 K for P = 17mW. In order to estimate $\Delta T$ with P = 50 mW, we compared the hysteresis loops at $T_{cryo}$ = 300 and 320 K with P = 0 mW to the hysteresis loops recorded at negative delay for $T_{cryo}$ = 80K with P = 50 mW (figure 3). The same signs of the hysteresis loops indicate that these measurements were performed above $T_{comp}$. Moreover, we notice that the coercive field is smaller for P = 50mW and $T_{cryo}$ = 80K compared to P = 0 mW and $T_{cryo}$ = 300 and 320 K. According to the thermal variation of the saturation field above $T_{comp}$ (figure 1), we estimated that $T_{cryo} + \Delta T$ is above 320K for P = 50mW and $T_{cryo}$ = 80K. The shape of the hysteresis loop at P = 50mW and $T_{cryo}$ = 80K also indicates that we are close to the temperature of spin reorientation transition [43]. As a consequence, we estimated $\Delta T$ > 250 K for P = 50 mW. Our procedure to estimate $\Delta T$ results in significant error bars on T\* (figure 5). Nevertheless, we estimated that we performed the time-resolved experiments at **T\*** = 350 K, 400 K and 540 K (figure 2, table 1). The measurements were carried out above $T_{comp}$ for the configurations 1 and 2 (figure 2a and b) and below $T_{comp}$ for the configurations 3 (figure 2c) [27].

## 3. *Experimental results and discussion*

The normalized transient XMCD signals recorded at the Co $L_3$ and Dy $M_5$ edges for $T^* =$ 350, 400 and 540 K are displayed in figure 4a, b and c respectively. At $T^*$=540K (figure 4c), the maximum demagnetization of the Co sublattice is reached while the demagnetization of the Dy sublattice has barely started as reported in a large number of element- and time-resolved experiments for different RE-TM alloys [21 - 27]. In contrast, at $T^* = 350$K (figure 4a), the magnetization of the Dy sublattice is close to its minimum value while the minimum magnetization of the Co sublattice is reached. The tr-XMCD curves at the Co $L_3$ edges were fitted with two exponential functions (respectively the demagnetization and the magnetization recovery) convoluted with a Gaussian function which account for the experimental time resolution (130 fs) [24, 44, 45]. It is worth noticing that we have imposed a lower limit at 130 fs for the characteristic demagnetization times during the fitting procedure. It means that the actual $\tau_{Co}$ is possibly below the experimental time resolution for $T^* = 400$ K and 540K. The tr-XMCD curves at the Dy $M_5$ edge were adjusted by a single exponential decay convoluted by a Gaussian function since we did not observe any recovery on the recorded time range. We have extracted the characteristic demagnetization times ($\tau$) from these fits as well as their error bars, which correspond to one standard deviation (table 1). The dependence of the characteristic demagnetization times on temperature for both sublattices are displayed in figure 5. We also report the characteristic demagnetization times for the Co and Dy sublattices in various CoDy alloys measured by Ferté et al (in $Co_{80}Dy_{20}$ and $Co_{78}Dy_{22}$) [27] and Radu et al (in $Co_{83}Dy_{17}$) [25] in figure 6. Ferté et al have explicitly given the numerical values for $T_{cryo}$, $T_{Curie}$ and $\Delta T$ [27] and therefore we have derived $T^* = 430$ and 450K for the $Co_{80}Dy_{20}$ and $Co_{78}Dy_{22}$ alloys respectively. Radu et al have performed their measurements at $T = 100$K without considering any DC heating. Therefore, we assume $\Delta T = 0$K and we include an extended error bar for this data. It is worth noticing that including a temperature elevation of $\Delta T$ ~100-200 K (typical values estimated for the pump-probe experiments on thin films) will not change our main message. Thus, we derived $T^*$=930K since $T_{Curie} = 1030$K for their $Co_{83}Dy_{17}$ alloys [43].

*Table 1: Characteristic demagnetization times extracted from the fit functions as a function of temperature T\*. The cryostat temperature ($T_{cryo}$), the laser continuous heating (ΔT), the laser power and the X-ray absorption edges are also recalled.*

| T* (K) | $T_{Cryo}$ (K) | ΔT (K) | Laser power (mW) | Edge | Demagnetization time τ (fs) |
|---|---|---|---|---|---|
| 540 | 80 | 80 | 17 | Co $L_3$ | 130 ± 100 |
| 540 | 80 | 80 | 17 | Dy $M_5$ | 980 ± 200 |
| 400 | 220 | 80 | 17 | Co $L_3$ | 130 ± 60 |
| 400 | 220 | 80 | 17 | Dy $M_5$ | 570 ± 90 |
| 350 | 80 | > 250 | 50 | Co $L_3$ | 212 ± 25 |
| 350 | 80 | > 250 | 50 | Dy $M_5$ | 400 ± 100 |

In figure 5, we observe that $τ_{Co}$ is almost constant within the error bars between T* = 350 K and T* = 540K in spite of the various laser powers used, and thus the different demagnetization amplitudes, in line with previous work by Jal et al [46]. We also observe a clear decrease of $τ_{Dy}$ when T* decreases from 540K to 350K (figure 5). In pure Gd layers, the characteristic demagnetization time $τ_{Gd}$ related to the (5d, 6s) magnetic order increases when the laser power is increased [13, 16]. The concomitant quenching of itinerant (5d, 6s) and localized 4f magnetic order in pure RE layers [12 - 16, 18, 47] suggests that such an increase is also expected for the 4f spins. However, such a behavior is not observed in our CoDy alloy since the shorter demagnetization time ($τ_{Dy}$) is obtained for the larger laser power (P = 50mW) and thus also for the larger demagnetization amplitude. Therefore, we can rule out the distinct laser power as the origin of the measured variation of $τ_{Dy}$ with temperature. Gang et al. have reported a decrease of the Ni 3d characteristic demagnetization time in NiPd ferromagnetic alloys when the Curie temperature is reduced (and thus T*) by increasing the Pd concentration [48]. They have attributed such feature to an increase of the spin-flip scattering probability [2] with Pd content, caused by its larger spin-orbit coupling compared to pure Ni. In our case, this explanation does not hold since we have studied a single alloy composition. According to the LLB calculations in ferrimagnets [32], by increasing the temperature, we would expect to go from a situation in which $τ_{TM} < τ_{RE}$ at low temperatures (T << $T_{Curie}$) to a situation in which $τ_{TM} \sim τ_{RE}$ at higher temperatures (T > 0.8 $T_{Curie}$). The transition between these dynamical regimes requires that $\tau_{Dy}$ decreases and/or $\tau_{Co}$ increases in the intermediate temperature range. Our experimental findings are thus consistent with this LLB prediction but call for further experiments at even higher temperatures, especially in the vicinity of $T_{Curie}$ to challenge further their predictions.

Two qualitative descriptions can be proposed to explain the different temperature dependent evolution of $\tau_{Co}$ and $\tau_{Dy}$. The first one considers that different microscopic mechanisms are supposed to be responsible for the laser induced ultrafast quenching of Co 3d and Dy 4f magnetic orders. Indeed, the dynamics of TM 3d spins is presumably caused by spin-flip scattering [49 - 51] and superdiffusive spin transport [3, 52] while the dynamics of the RE 4f spins is claimed to be related to spin-waves [53 - 55]. The second one considers that the Co 3d spin dynamics is mainly governed by the direct ferromagnetic Co – Co exchange coupling while the Dy 4f spin dynamics is mainly governed by the indirect antiferromagnetic Co – Dy exchange coupling. Recent measurements have shown that antiferromagnetic and ferromagnetic materials exhibit distinct laser induced ultrafast dynamics [56].

Finally, in figure 6, we compare our experimental results with existing data from the literature [31, 33] by plotting $\tau_{Co}$ and $\tau_{Dy}$ as a function of T* for different alloy compositions. We observe that $\tau_{Co}$ and $\tau_{Dy}$ are both constant (within error bars) from T* = 930 K to T* = 540 K. From T* = 540 K to T* = 350 K, we observe that $\tau_{Co}$ is constant while $\tau_{Dy}$ decreases, confirming our experimental results from figure 5. It is also very interesting to note that the $\tau_{Dy}$ extracted from the work of Ferté et al. are consistent with our present results for similar T* although in their case the demagnetization amplitudes were larger than 80%. It suggests that $\tau_{Dy}$ does not dependent on the demagnetization amplitude but mainly on T* = $T_{Curie}$ – T.

### 4. Conclusions

We have investigated the laser induced ultrafast dynamics of Co 3d and Dy 4f spins in a ferrimagnetic CoDy alloy as a function of temperature by element- and time-resolved XMCD. We have revealed striking differences between the Co 3d and Dy 4f spin dynamics when the temperature is varying from 160 K to 350K. The characteristic demagnetization time of the Dy4f sublattice decreases while it is almost constant for the Co3d sublattice. This experimental findings sustain some of the predictions of the LLB model, namely that the characteristic demagnetization time of the RE sublattice should be smaller than that of the TM sublattice at high temperature. Our experimental results also confirm that the characteristic demagnetization time of the Co sublattice does not depend on the composition, on the demagnetization amplitude nor on the temperature in CoDy alloys as reported by Jal et al [46]. Finally, our data set, amended with data extracted from literature, suggest that the characteristic demagnetization time of the Dy sublattice is determined by $T^* = T_{Curie} - T$ and does not depend on the demagnetization amplitude or alloy composition. Our work calls for further experimental investigations at elevated temperatures to challenge the predictions of the LLB model for spin dynamics in RE-TM alloys, especially in the vicinity of $T_{Curie}$. Such experimental confirmation would link the characteristics of laser induced ultrafast dynamics and the static magnetic properties of ferrimagnetic alloys [32]. We hope this work will motivate further experimental investigation at elevated temperatures as well as development of the LLB model to ferrimagnetic alloys [57].

***Figures*:**

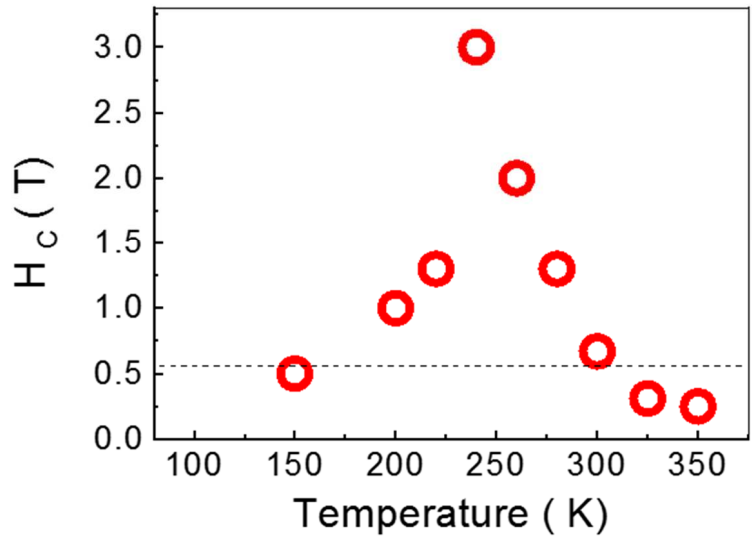

*Figure 1: Dependence of the coercive field ($H_C$) on the temperature for a $Co_{80}Dy_{20}$ alloy measured by means of SQUID magnetometry (red empty circles). $H_C$ diverges in the vicinity of T ~ 250K defining the temperature of magnetic compensation ($T_{comp}$ ~ 250K). The horizontal dotted line corresponds to H = 0.55 T which is the maximum magnetic field available on the femtoslicing end-station [37]. The $Co_{80}Dy_{20}$ alloy layer used for SQUID measurements was deposited simultaneously with the one used for the time-resolved experiments deposited on transparent SiN membrane.*

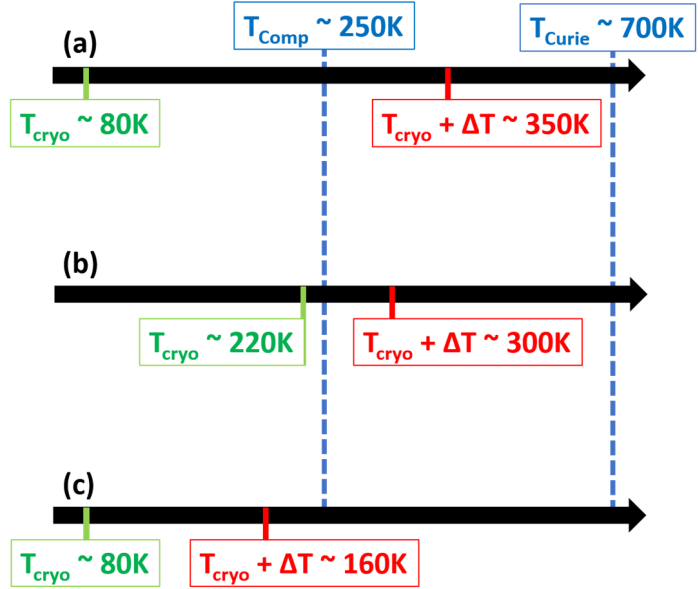

*Figure 2: Sketch of the experimental conditions to achieve our different relative temperatures T\* = $T_{Curie}$ - ($T_{cryo}$-ΔT) = 350K (a), 400K (b) and 540K (c).*

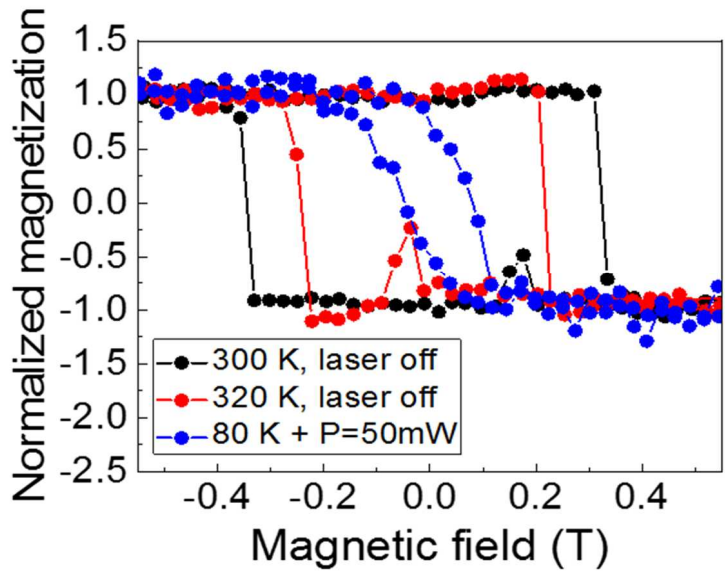

*Figure 3: Hysteresis loops recorded by monitoring the X-ray transmission at the Dy $M_5$ absorption edge as a function of the magnetic field. The experimental configurations were P = 0mW and $T_{cryo}$ = 300K (black circles), P = 0mW and $T_{cryo}$ = 320K (red circles) and P = 50mW and $T_{cryo}$ = 80K (black circles).*

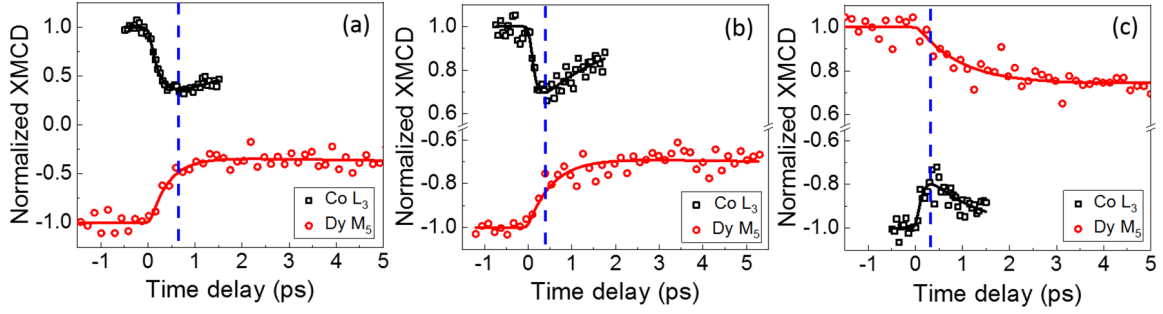

*Figure 4*: Transient XMCD at the Co $L_3$ (black squares) and Dy $M_5$ (red circles) edges as a function of the pump – probe delay measured at T* = 350 K (a), 400 K (b) and 540 K (c). The solid lines are the fitting functions. The vertical blue dotted lines denote the delay at which the magnetization of the Co sublattices reaches the minimal value.

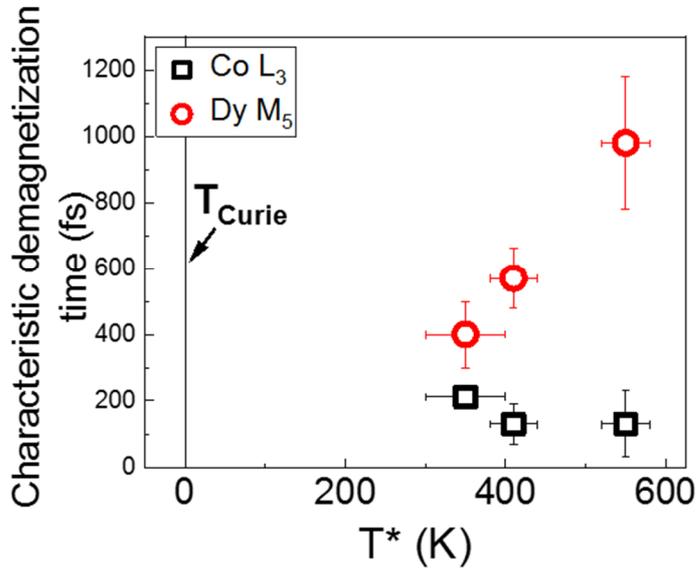

*Figure 5:* Characteristic demagnetization times for Co (black squares) and Dy (red circles) sublattices as a function of T*. The error bars on the characteristic demagnetization times is given by the standard deviation from the fitting function. The error bars on the temperature were experimentally estimated (details in the text).

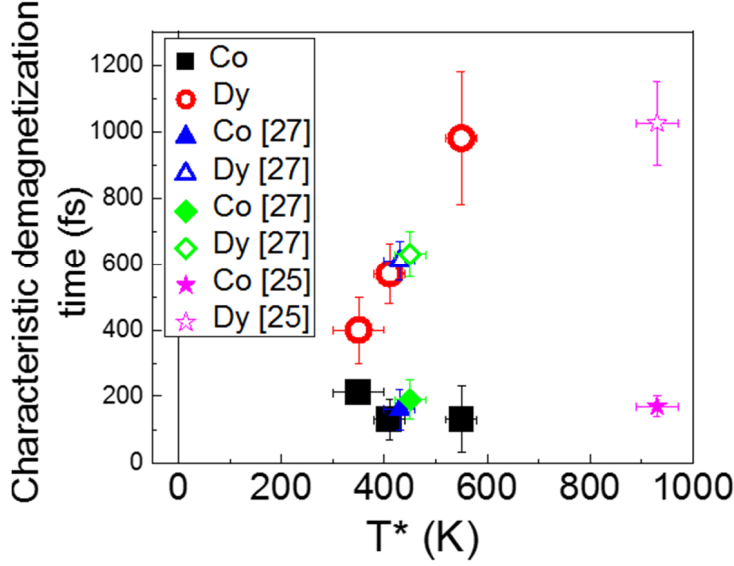

*Figure 6: Characteristic demagnetization times for Co (filled symbols) and Dy (open symbols) sublattices as a function of T\*. We report our experimental results (black squares and red circles) superposed to published data extracted from element and time-resolved experiments performed on $Co_{80}Dy_{20}$ (filled and empty blue triangles) [27], $Co_{78}Dy_{22}$ (filled and empty green lozenges) [27] and $Co_{83}Dy_{17}$ (filled and empty magenta stars) [25].*

*Acknowledgments*:

We thank HZB for the allocation of synchrotron radiation beamtime. We are indebted for the scientific and technical support given by N. Pontius, Ch. Schüßler-Langeheine and R. Mitzner at the slicing facility at the BESSY II storage ring. We thank D. Gupta for a careful reading and her suggestions to improve the manuscript. This project has received funding from the European Union's Horizon 2020 research and innovation programme under grant agreement No 730872. The authors are grateful for financial support received from the following agencies: the French "Agence National de la Recherche" via Project No. ANR-11-LABX-0058_NIE and Project EQUIPEX UNION No. ANR-10-EQPX-52, the CNRS-PICS program, the EU Contract Integrated Infrastructure Initiative I3 in FP6 Project No.R II 3CT-2004-50600008. Experiments were carried out on the IJL Project TUBE-Davms equipment funded by FEDER (EU), PIA (Programme Investissemnet d'Avenir), Region Grand Est, Metropole Grand Nancy, and ICEEL.

The authors have no competing interests to declare